\documentclass[preprint,nofootinbib,superscriptaddress]{revtex4-1}
\usepackage{xcolor}
\usepackage{verbatim}
\usepackage{amsmath}
\usepackage{amssymb}
\usepackage{graphicx}
\usepackage[unicode=true,pdfusetitle,
 bookmarks=true,bookmarksnumbered=false,bookmarksopen=false,
 breaklinks=true,pdfborder={0 0 1},backref=false,colorlinks=false]
 {hyperref}

\makeatletter

\providecommand{\tabularnewline}{\\}

\@ifundefined{date}{}{\date{}}
\@ifundefined{showcaptionsetup}{}{%
 \PassOptionsToPackage{caption=false}{subfig}}
\usepackage{subfig}
\makeatother

\begin{document}
	
\count\footins = 1000 
\title{Exploring dark sector parameters in light of neutron star temperatures}
\author{Guey-Lin Lin}
\email{glin@nycu.edu.tw}
\affiliation{Institute of Physics, National Yang Ming Chiao Tung University, Hsinchu
	30010, Taiwan}

\author{Yen-Hsun Lin}
\email{yenhsun@gate.sinica.edu.tw}

\affiliation{Institute of Physics, National Yang Ming Chiao Tung University, Hsinchu
30010, Taiwan}
\affiliation{Institute of Physics, Academia Sinica, Taipei
11529, Taiwan}

\begin{abstract}
Using neutron stars (NS) as a dark matter (DM) probe has gained  broad
attention recently, either from heating due to DM annihilation or
its stability under the presence of DM. In this work, we investigate
spin-$1/2$ fermionic DM $\chi$ charged under the $U(1)_{X}$
in the dark sector. The massive gauge boson $V$ of
$U(1)_{X}$ gauge group can be produced in NS via DM annihilation. The produced gauge boson can decay
into Standard Model (SM) particles before it exits the NS, despite its
tiny couplings to SM particles. Thus, we perform a systematic study on $\chi\bar{\chi}\to2V\to4{\rm SM}$
as a new heating mechanism for NS in addition to $\chi\bar{\chi}\to2{\rm SM}$
and kinetic heating from DM-baryon scattering. The self-trapping due
to $\chi V$ scattering is also considered. 
We assume the general framework that both kinetic and mass mixing terms between $V$ and SM gauge bosons are present. This allows both vector and axial-vector couplings between $V$ and SM fermions even for $m_V\ll m_Z$. 
Notably, the contribution from axial-vector coupling is not negligible
when particles scatter relativistically. 
We point out that the above approaches to DM-induced NS heating are not yet adopted
in recent analyses. Detectabilities of the aforementioned
effects to the NS surface temperature by the future telescopes are
discussed as well.
\end{abstract}
\maketitle

\section{Introduction}


It has been widely accepted that one-fifth of the total energy of
the Universe consists of dark matter (DM). Though multidisciplinary
strategies are employed to identify its essence, either from direct
\cite{Aad:2015zva,Abdallah:2015ter,Aalbers:2016jon,Akerib:2016vxi,Amole:2017dex,Akerib:2017kat,Aprile:2017iyp,Aprile:2018dbl,Aprile:2019xxb,Aprile:2019jmx} or indirect detections \cite{Aartsen:2014oha,Choi:2015ara,Aartsen:2016zhm,Aguilar:2015ctt,TheFermi-LAT:2017vmf,Ambrosi:2017wek,Beck:2021xsv}, the nature of DM remains a puzzle. The approach of using neutron stars (NS) as the DM probe has
been proposed from the heating effect due to DM  \cite{Kouvaris:2007ay,deLavallaz:2010wp,Kouvaris:2010vv,Baryakhtar:2017dbj,Raj:2017wrv,Chen:2018ohx,Bell:2018pkk,Acevedo:2019agu,Joglekar:2019vzy,Keung:2020teb,Dasgupta:2020dik,Garani:2020wge,Joglekar:2020liw},
the NS instability caused by DM gravitational collapse \cite{Kouvaris:2010jy,Leung:2011zz,Kouvaris:2011gb,McDermott:2011jp,Guver:2012ba,Bramante:2013hn,Bramante:2013nma,Kouvaris:2013kra,Gresham:2018rqo,Grinstein:2018ptl,Garani:2018kkd,Lin:2020zmm,Dasgupta:2020dik}
and gravitation wave emitted from the merger of binary NS admixed with DM \cite{Nelson:2018xtr,Ellis:2018bkr,Bauswein:2020kor}.
Novel ways of constraining long-lived particles through the NS in the Milky Way was also investigated recently \cite{Leane:2021ihh}.
In addition, DM self-interaction naturally arises in various phenomenological models and was proposed to resolve
many issues in the small-scale structure, e.g.~core-cusp, missing
satellite, too-big-to-fail and diverse galactic rotation curve, see
Ref.~\cite{Tulin:2017ara} for a review. Current astrophysical observations constrain
DM self-interaction cross section $\sigma_{\chi\chi}$ in the range \cite{Randall:2007ph,Walker:2011zu,BoylanKolchin:2011de,BoylanKolchin:2011dk,Elbert:2014bma}
\begin{equation}
0.1\,{\rm cm}^{2}\,{\rm g}^{-1}\leq\sigma_{\chi\chi}/m_{\chi}\leq10\,{\rm cm}^{2}\,{\rm g}^{-1}\label{eq:sidm}
\end{equation}
where $m_{\chi}$ is the DM mass.

DM self-interaction can
be understood phenomenologically as an exchange of the $U(1)_X$ gauge boson
$V$ 
between dark matter particles.
Assuming DM $\chi$ is a spin-$1/2$ fermion carrying $U(1)_{X}$ dark charge $g_{d}$, its interaction with $V$ is given by the Lagrangian
\begin{equation}
\mathcal{L}_{\rm DM}=
i\bar{\chi}\gamma^{\mu}(\partial_{\mu}-ig_dV_{\mu})\chi-m_{\chi}\bar{\chi}\chi \label{eq:Ldm}
\end{equation}
where the associated DM self-interaction cross section $\sigma_{\chi\chi}$ can be calculated from $\mathcal{L}_{\rm DM}$ and constrained by Eq.~(\ref{eq:sidm}).
Here we consider the scenario of symmetric dark matter~\cite{Lin:2011gj} where the numbers $\chi$ and $\bar{\chi}$ are identical. It has been proposed that the vector boson $V$ in the dark sector can
mix with SM photons and $Z$ bosons through kinetic~\cite{Holdom:1985ag,Galison:1983pa,Foot:2004pa,Feldman:2006wd,ArkaniHamed:2008qn,Pospelov:2008jd} and mass mixing terms~\cite{Babu:1997st,Davoudiasl:2012ag,Davoudiasl:2013aya}; 
the latter generally arise from extended Higgs sectors. Ref.~\cite{Davoudiasl:2012ag} provides an example by introducing two Higgs doublet $\Phi_1$, $\Phi_2$, with $\Phi_1$ coupled to SM fermions, and one scalar Higgs singlet $\phi$. Both $\Phi_2$ and $\phi$ carry the dark charge $g_d$ and the mixing among $\Phi_1$, $\Phi_2$ and $\phi$ are neglected for simplicity. Hence neither $\Phi_2$ nor $\phi$ couples to SM fermions.
The vacuum expectation values of Higgs scalars give mass terms for $Z$, $V$, and their mixing, which, together with kinetic mixing terms, are given by 
\begin{eqnarray}
\mathcal{L}_{\rm gauge}&=&-\frac{1}{4}B_{\mu\nu}B^{\mu\nu}+\frac{1}{2}\frac{\varepsilon_\gamma}{\cos\theta_W}B_{\mu\nu}V^{\mu\nu}-\frac{1}{4}V_{\mu\nu}V^{\mu\nu},\label{eq:L_gauge} \\
\mathcal{L}_{\rm mass}&=&\frac{1}{2}m_Z^2Z_{\mu}Z^{\mu}-\varepsilon_Z m_Z^2Z_{\mu}V^{\mu}+\frac{1}{2}m_V^2 V_{\mu}V^{\mu},
\label{eq:L_mass}
\end{eqnarray}
where $B^{\mu\nu}\equiv \partial_{\mu}B_{\nu}-\partial_{\nu}B_{\mu}$ is the $U(1)_Y$ field strength in SM while
$\varepsilon_{\gamma}$ and $\varepsilon_{Z}$ are the
kinetic and $V-Z$ mass mixing parameters respectively. 
It is important to note that we do not invoke spontaneous symmetry breaking for generating DM mass as can be seen from Eq.~\eqref{eq:Ldm} \cite{Bell:2016uhg}. Since none of the scalar fields mentioned above couple to $\chi$, the only mediator between the dark and visible sector is the dark boson $V$. Explicitly speaking,
the electromagnetic (EM) and neutral-current (NC) interactions between
$V$ and SM fermions $f$ resulting from mixing terms in Eqs.~(\ref{eq:L_gauge}) and (\ref{eq:L_mass}) are given by 
\begin{equation}
\mathcal{L}_{{\rm DS-SM}}=\left(\varepsilon_{\gamma}eJ^{\rm EM}_{\mu}+\tilde{\varepsilon}_{Z}\frac{g_{2}}{\cos \theta_{W}}J^{\rm NC}_{\mu}\right)V^{\mu}
\label{DS-SM}
\end{equation}
where $g_2$ is the $SU(2)_L$ coupling and $J_\mu^{{\rm EM}}$ and $J_\mu^{{\rm NC}}$ are the SM electromagnetic and neutral currents, 
respectively. The coefficient $\tilde{\varepsilon}_Z$ is a linear combination of two mixing parameters and it reduces to $\varepsilon_Z$ for $m_V\ll m_Z$. Its general expression is given in Appendix~\ref{sec:XN_int}.  

In this paper, we examine the effect of DM heating due to the above phenomenological
setup for a nearby three giga-year-old (Gyr-old) and isolated NS. The
associated temperature is around $100\,{\rm K}$ according to the
standard cooling mechanism if there is no other heating source.
Therefore, any temperature deviation from this benchmark value can
be potentially due to DM annihilation in the star. DM annihilation channels in this regard include not only $\chi\bar{\chi}\to f\bar{f}$ but also $\chi\bar{\chi}\to 2V$,
provided $m_V<m_\chi$ and the decay length of $V\to f\bar{f}$ is smaller than the radius of the star. Since $V$ couples to neutral scalar bosons to acquire its own mass, the annihilation process $\chi\bar{\chi}\to V^*\to V+s$ with $s$ one of the neutral scalar bosons is also possible for $m_V+m_s< 2m_{\chi}$. Such a process yields comparable heating effect to that given by $\chi\bar{\chi}\to 2V$ but involves an additional mass parameter $m_s$. For simplicity in our discussions, we shall not consider this kinematic region. We refer the readers to Refs.~\cite{Bell:2016fqf,Duerr:2016tmh} for the phenomenology of such an annihilation channel.

Searching the
nearby old and cold NS can improve our understanding about DM. The new
dynamics emerging from the above phenomenological setup will be discussed
in the following sections. For completeness, we also analyze the
signal to noise ratio (SNR) in the James Webb Space Telescope (JWST)
\cite{Gardner:2006ky}. Future telescopes such as European Extremely Large Telescope
(E-ELT) and Thirty-Meter Telescope (TMT) \cite{Skidmore:2015lga} will constrain
DM properties with unprecedented sensitivities. In the following sections, we employ the NS mass $M_{0}=1.4M_{\odot}$ and 
and the radius $R_{0}=12\,{\rm km}$. We also replace $g_{d}$ with
$\alpha_{\chi}=g_{d}^{2}/4\pi$ and all equations are expressed in
terms of natural units $\hbar=c=k_{B}=1$.

\section{DM capture and NS temperature}

When a NS swipes through space, the DM particles in the halo can
scatter with the baryons and leptons inside the star. Once DM loses
an appreciable fraction of kinetic energy, it will be gravitationally  
captured by the NS. This capture process has been investigated extensively
with contributions from neutrons, protons and leptons as well as relativistic
corrections included in Refs.~\cite{Bell:2020jou,Bell:2020lmm}. In this paper, only neutron
contribution to the capture rate $C_{c}$ is considered. Contributions
from other particle species are ignored due to their small yields.
The DM number $N_{\chi}$ in the
star satisfies the differential equation
\begin{equation}
\frac{dN_{\chi}}{dt}=C_{c}-C_{a}N_{\chi}N_{\bar{\chi}},
\label{eq:dm_number}
\end{equation}
while the anti-DM number $N_{\bar{\chi}}$ evolves according to
\begin{equation}
\frac{dN_{\bar{\chi}}}{dt}=C_{c}-C_{a}N_{\bar{\chi}}N_{\chi}.
\label{eq:antidm_number}
\end{equation}
Here $C_{a}$ is the DM annihilation rate.
Both coefficients $C_c$ and $C_a$ are well studied and the expressions can be found in Refs.~\cite{Bell:2020jou,Bell:2020lmm,Chen:2018ohx} and references therein.
We do not reproduce here.
Thus, the exact solutions to Eqs.~(\ref{eq:dm_number}) and (\ref{eq:antidm_number}) are obtained
\begin{equation}
    N_{\chi}=N_{\bar{\chi}}=C_{c}\tau_{{\rm eq}}\tanh\left(\frac{t}{\tau_{{\rm eq}}}\right)\label{eq:Nx}
\end{equation}
where $\tau_{{\rm eq}}=1/\sqrt{C_{c}C_{a}}$ is the equilibrium timescale.
Once $t>\tau_{{\rm eq}}$, $dN_{\chi}/dt=0$, and 
$N_{\chi}(t>\tau_{{\rm eq}})=\sqrt{C_{c}/C_{a}}$ according to Eq.~(\ref{eq:dm_number}). The total annihilation
rate at this stage only depends on the capture rate since $\Gamma_{a}=C_{a}N_{\chi}N_{\bar{\chi}}=C_{c}$.\footnote{The $C_a$ depends on the annihilation cross section $\langle\sigma v\rangle$ explicitly but its effect only appears in the total annihilation rate $\Gamma_A$ through $\tanh(t/\tau_{\rm eq})$ as $\tau_{\rm eq}^2\propto 1/\langle\sigma v\rangle$. No matter what the value of $\langle\sigma v\rangle$ is, one always has $\tanh(t/\tau_{\rm eq})\approx 1$ as long as the NS age is greater than the equilibrium time scale. In this case the total annihilation rate $\Gamma_A$ no longer depends on $\langle\sigma v\rangle$, but rather it is solely determined by the capture rate $C_c$. We have carefully examined that even
$\langle\sigma v\rangle$ is an order of magnitude smaller than the thermal relic one given in
Ref.~\cite{Steigman:2012nb}, the NS can still attain equilibrium at $t\sim{\rm Gyr}$ and have no sensitivity on $\langle\sigma v\rangle$ anymore. The typical timescale for $\tau_{\rm eq}$ is about hundreds to thousands of years for $(\langle\sigma v\rangle,\sigma_{\chi n})=(6\times 10^{-26}\,{\rm cm^3\,s^{-1}},10^{-47}\,{\rm cm^2})$ for $m_\chi$ around MeV to TeV. }
Note that $C_{c}$ depends on $\sigma_{\chi n}$ and $\sigma_{\chi n}\leq\sigma_{\chi n}^{{\rm geom}}\approx10^{-44}\,{\rm cm}^{2}$
where $\sigma_{\chi n}^{{\rm geom}}$ is the geometric cross section.
In principle, the maximum capture rate is determined by $C_{c}(\sigma_{\chi n}^{{\rm geom}})=C_{c}^{{\rm geom}}$.
Besides, when DM falls into the NS surface, it is accelerated up to
$0.3c-0.5c$. The nonrelativistic (NR) limit for calculating $\sigma_{\chi n}$ is not
applicable. Furthermore one has to consider contributions from axial-vector coupling due to $V-Z$ mass mixing
given by $\mathcal{L}_{\rm mass}$. We have thoroughly included these effects. A brief discussion
on how to compute $\sigma_{\chi n}$ in terms of relativistic kinematics
is given in Appendix~\ref{sec:XN_int}.

NS is known to suffer from eternal cooling due to neutrino
and photon emissions. Without extra energy injection, the
NS temperature drops until it releases all its heat. However, if SM particles are produced due
to DM annihilation in the star, these particles can become a heat
source and potentially prevent the star from inevitable cooling.
Therefore, the evolution of NS interior temperature $T_{b}$ is governed 
by the equation
\begin{equation}
\frac{dT_{b}}{dt}=\frac{-\epsilon_{\nu}-\epsilon_{\gamma}+\epsilon_{\chi}}{c_{V}},\label{eq:dTdt}
\end{equation}
where $\epsilon_{\nu}\approx2.1\times10^{4}\,{\rm erg}\,{\rm cm}^{-3}\,{\rm s}^{-1}\,(T_{b}/10^{7}\,{\rm K})^{8}$
is the neutrino emissivity, $\epsilon_{\gamma}\approx1.8\times10^{14}\,{\rm erg}\,{\rm cm}^{-3}\,{\rm s}^{-1}\,(T_{b}/10^{8}\,{\rm K})^{2.2}$
is the photon emissivity, $\epsilon_{\chi}$ is the DM emissivity that is responsible for the heating from DM
annihilation, and $c_{V}$ the NS heat capacity \cite{Kouvaris:2007ay}. Additionally,
the surface temperature $T_s$ observed by a distant observer is related to $T_b$ by $T_{s}\approx8.7\times10^{5}\,{\rm K}\,(g_{s}/10^{14}\,{\rm cm}\,{\rm s}^{-1})^{1/4}(T_{b}/10^{8}\,{\rm K})^{0.55}$
where $g_{s}=GM/R^{2}\approx1.85\times10^{14}\,{\rm cm\,s}^{-2}$
accounts for the redshift correction from the surface gravity of the star.
It is also pointed out that when $T_{b}<3700\,{\rm K}$, there is
no distinction between $T_{b}$ and $T_{s}$ \cite{Chen:2018ohx}.

During each annihilation, a pair of DMs release $2m_{\chi}$ of energy
in a form of SM particles or dark bosons depending on which channels
are kinematically allowed. The total energy release rate by DM is
$\mathcal{E}_{\chi}=2m_{\chi}\Gamma_{a}\sum_{i}b_{i}$ where $b_{i}$
is the branching ratio of a specific channel, e.g.,~$e^{\pm},\mu^{\pm},\tau^{\pm}$
or $q\bar{q}$, and $\sum_{i}b_{i}\leq1$. Neutrino pair $\nu\bar{\nu}$
is also part of the annihilation channel in the presence of $V-Z$ mass mixing,
but it cannot contribute to the heating. In addition to the annihilation,
DM also loses kinetic energy $E_{k}$ to
the star through the capturing process. This has been realized as the kinetic
heating \cite{Baryakhtar:2017dbj} with the rate  $\mathcal{K}_{\chi}=C_{c}E_{k}=C_{c}m_{\chi}(\gamma-1)$,
where $\gamma=1/\sqrt{1-v^{2}}$ is the Lorentz factor.\footnote{Even if DM is not captured, 
energy deposition still occurs as long as $\chi n$ scattering can happen. On the other hand, the kinetic heating effect
from such uncaptured DM is relatively small and negligible in our calculation.
} Thus, DM emissivity $\epsilon_{\chi}$ is given by
\begin{equation}
\epsilon_{\chi}=\frac{\mathcal{E}_{\chi}+\mathcal{K}_{\chi}}{V},\label{eq:DM_heat_coeff}
\end{equation}
where $V$ is the NS volume.

\section{Decays of dark boson}
\begin{figure*}
	\begin{centering}
		\subfloat[\label{fig:V_decay}]{\begin{centering}
				\includegraphics[width=0.25\textwidth]{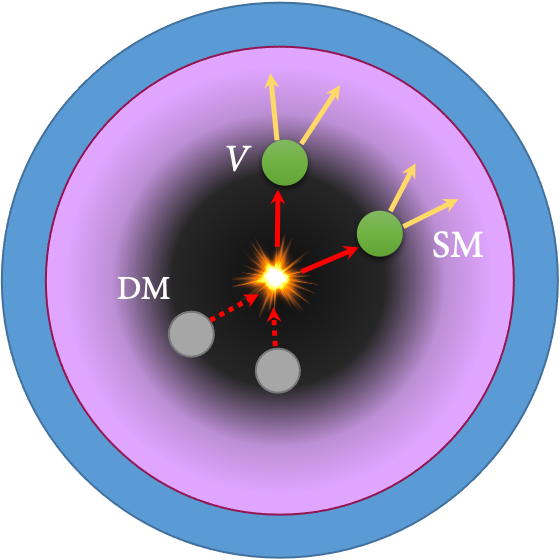}
				\par\end{centering}}\quad
		\subfloat[\label{fig:V_self_trap}]{\begin{centering}
				\includegraphics[width=0.25\textwidth]{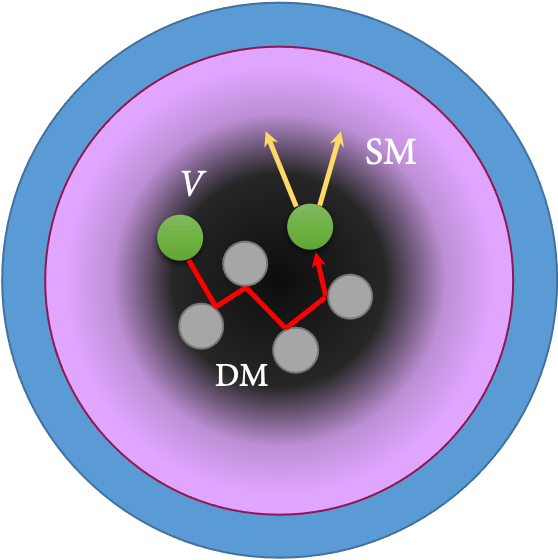}
				\par\end{centering}
		}
		\par\end{centering}
	\caption{DM heating from dark boson production $\chi\bar{\chi}\to2V$. 
	{\it Left}: $V$ decays into SM particles before it exits the star.
	{\it Right}: $V$ is self-trapped due to multiple $\chi V$ scatterings  and then decays. }
\end{figure*}
Here we discuss the case of $V$ produced by DM annihilation.
$V$ is usually produced in DM rich environment.
If $V$ can subsequently scatter off the surrounding DM multiple times,
it could lose energy and be self-trapped. It then decays promptly as shown in
Fig.~\ref{fig:V_self_trap}.
However, such self-trapping effect is in general inefficient since the $\chi V$ scattering length $\ell_{\chi V}$ is much larger than the
thermal radius $r_{\rm th}$. Hence the scattering rate is suppressed
and irrelevant to the heating. We show detailed discussions in Appendix~\ref{sec:V_in_NS}.
Another trapping is due to the scattering between $V$ and neutrons. On the other hand
the relevant cross section is further suppressed by the factor $\tilde{\varepsilon}_Z^4$ and
the scattering length is expected to be much larger than the NS radius.
It is safe to omit this effect in our calculation as well.

However, $V$ can decay into other SM particles before it propagates
to the surface as long as the decay length $\ell_{{\rm dec}}$ is
shorter than $R_{0}$ (see Fig.~\ref{fig:V_decay}). The decay length is given by $\ell_{{\rm dec}}=v\gamma\tau_{{\rm dec}}$
with $v\equiv \sqrt{1-m_{V}^{2}/m_{\chi}^{2}}$ the velocity of $V$, and $\tau_{{\rm dec}}\equiv \Gamma_{V}^{-1}$ the lifetime of $V$ at rest where $\Gamma_V$ is the total decay width.
Since $V$ is produced on shell,
we do not consider $V$ decaying back to $\chi$ due to $m_{V}<m_{\chi}$.
The probability for $V$ to convert into SM particles after a propagation
distance $r$ is
\begin{equation}
     F=1-e^{-r/\ell_{{\rm dec}}}. \label{eq:F}
\end{equation}
We took $r=R_{0}$ in the calculation. However, if neutrino is the decay product, it cannot be considered as the heating source and must
be subtracted. By examining the numerical results for $F$, we found
that $V$ can decay before it exits the star in most of our parameter
space of interest. This implies that $\chi\bar{\chi}\to2V$
also plays an important role in NS heating (see Appendix~\ref{sec:V_in_NS} for details). Generally speaking, NS
contain muons and electrons that are degenerate. To enable the decay $V\to \ell\bar{\ell}$ with $\ell$ corresponding to either the electron or the muon, ${V}$ should not only be heavier than $2m_{\ell}$, it also has to be energetic enough so that the
kinetic energy of $\ell$ exceeds the chemical
potential of $\ell$ for preventing the Pauli blocking effect. This condition has been implemented in
our study.

Given the information in this section, we summarize that even when
$\chi\bar{\chi}\to2V$ dominates the annihilation channel for $m_{V}<m_{\chi}$,
the heating effect is still efficient due to $V$ decays. However, the self-trapping
is generally unimportant due to $\ell_{\chi V}\gg r_{{\rm th}}$ in
this paper.

\section{Implication of DM on NS temperature\label{sec:implication}}

\begin{figure*}
\begin{centering}
\includegraphics[width=0.95\textwidth]{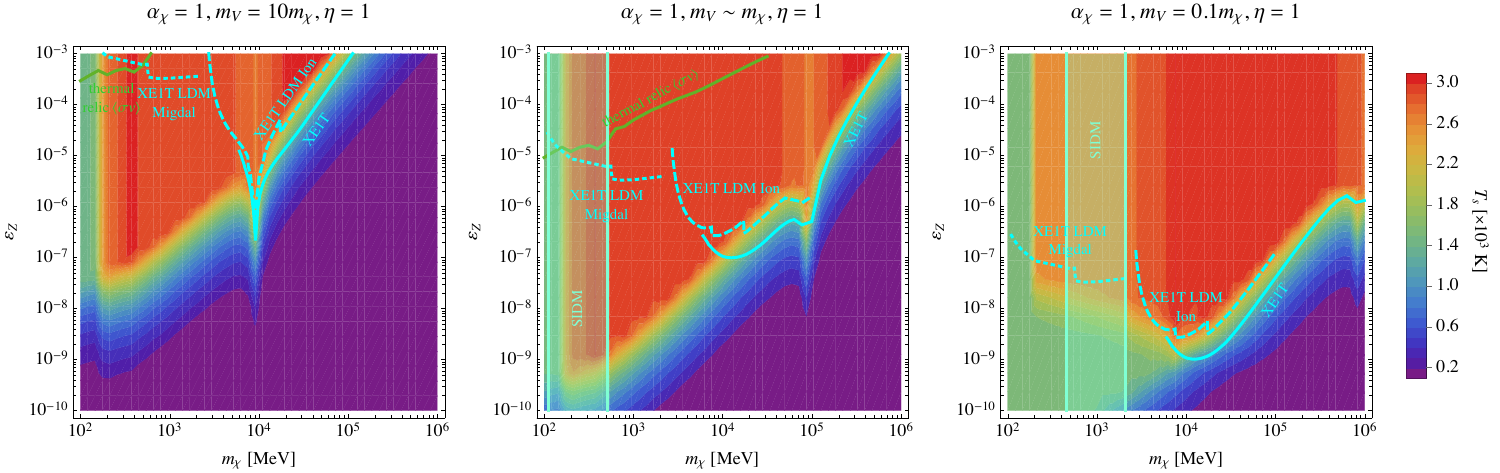}
\par\end{centering}
\begin{centering}
\includegraphics[width=0.95\textwidth]{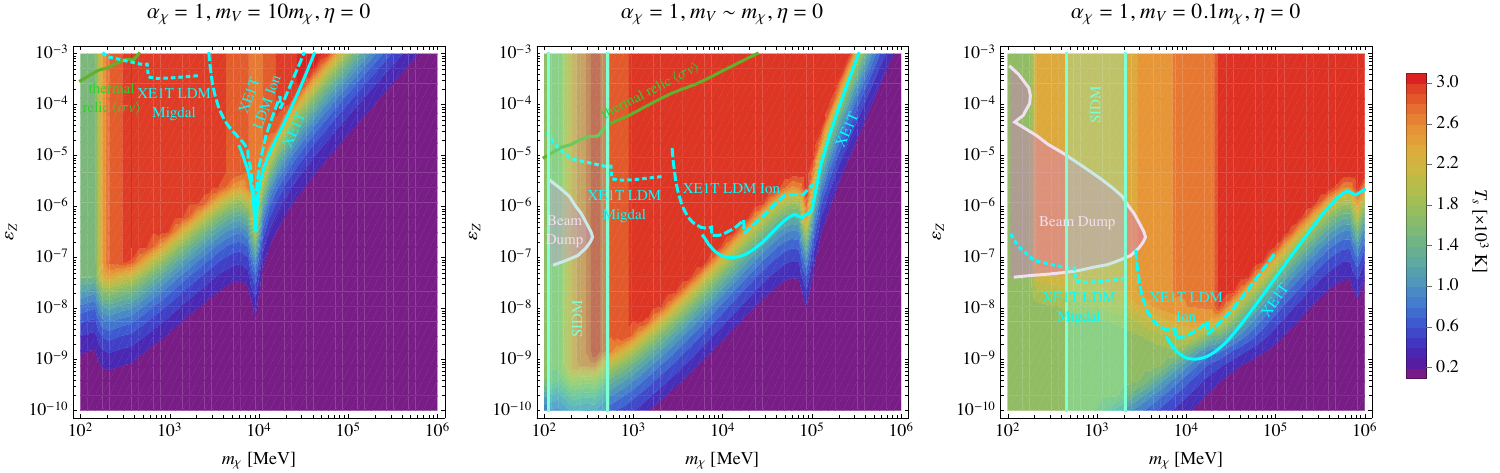}
\par\end{centering}
\begin{centering}
\includegraphics[width=0.95\textwidth]{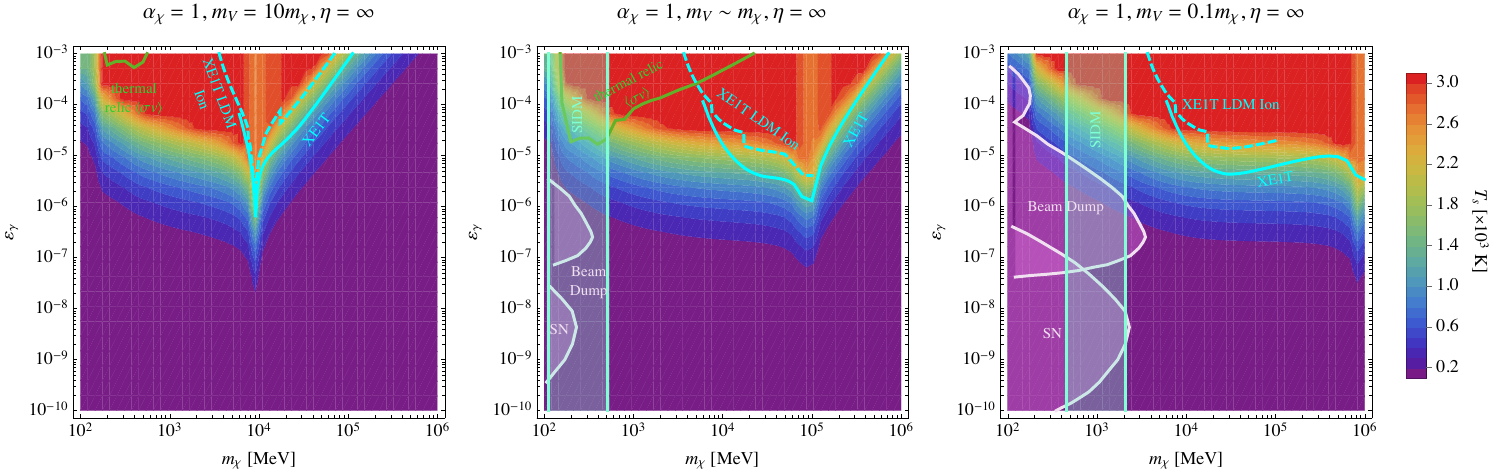}
\par\end{centering}
\caption{\label{fig:Ts_ax1}NS surface temperature $T_{s}$ in the $m_{\chi}-\varepsilon$
plane. We took the age of NS is 3 Gyrs and the lowest $T_{s}=100\,{\rm K}$
without DM heating. All figures have $\alpha_{\chi}=1$ and $\eta=\varepsilon_{\gamma}/\varepsilon_{Z}$.
From top to bottom, $\eta=1,0$, and $\infty$.
From left to right, $m_{V}/m_{\chi}=10,1,0.1$.
Various constraints
from XENON1T \cite{Aprile:2018dbl}, XENON LDM \cite{Aprile:2019xxb,Aprile:2019jmx}, SIDM \cite{Randall:2007ph,Walker:2011zu,BoylanKolchin:2011de,BoylanKolchin:2011dk,Elbert:2014bma}, SN1987A
\cite{Sung:2019xie,Sung:2021swd}, beam dump experiments \cite{Riordan:1987aw,Bross:1989mp,Abdullah:2018ykz} and the parameter curve rendering
thermal relic cross section are shown as well.}
\end{figure*}

\begin{figure*}
\begin{centering}
\includegraphics[width=0.95\textwidth]{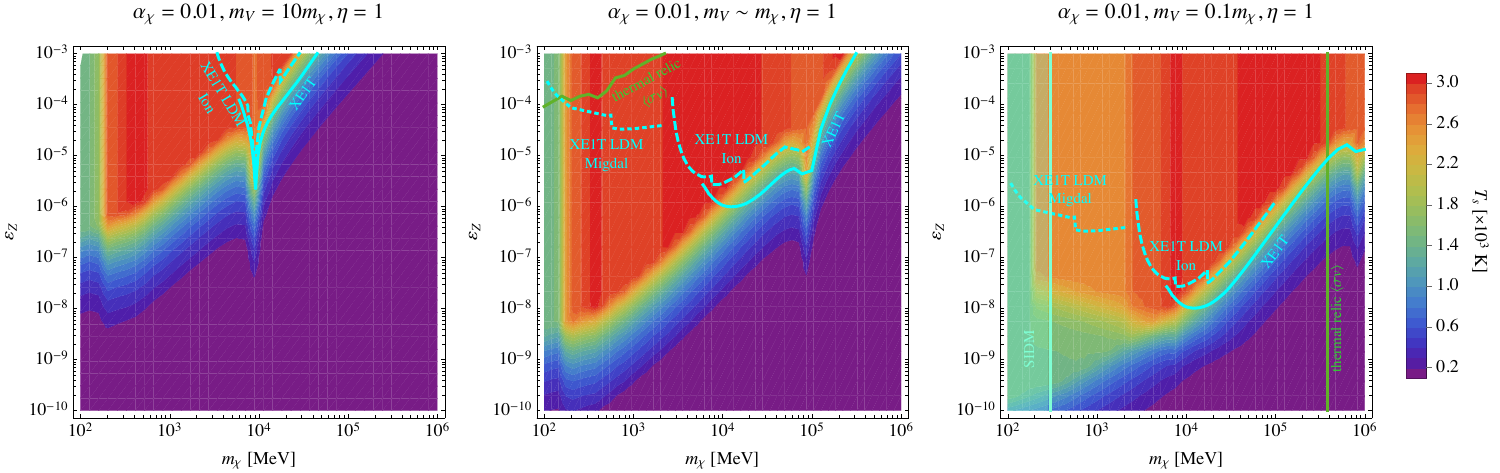}
\par\end{centering}
\begin{centering}
\includegraphics[width=0.95\textwidth]{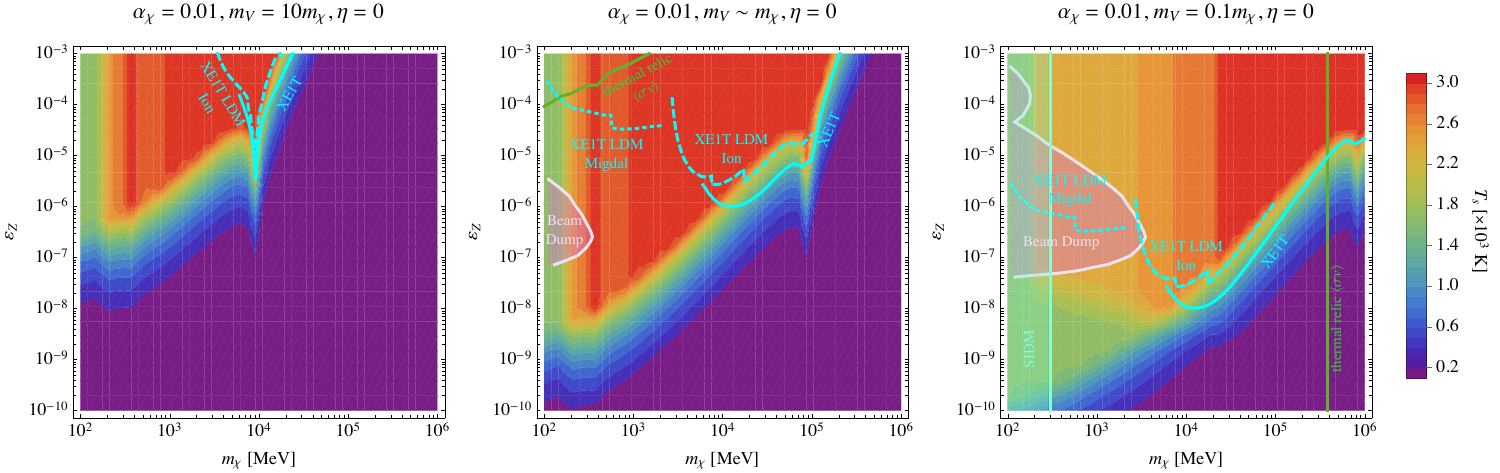}
\par\end{centering}
\begin{centering}
\includegraphics[width=0.95\textwidth]{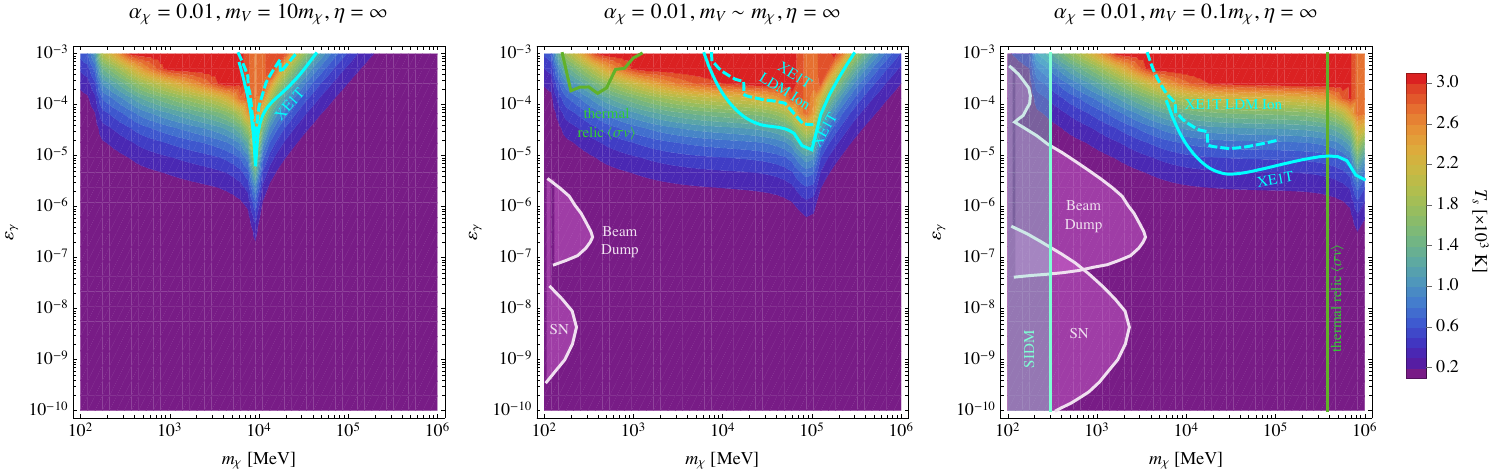}
\par\end{centering}
\caption{\label{fig:Ts_ax001}The same as Fig.~\ref{fig:Ts_ax1} except $\alpha_{\chi}=0.01$.}
\end{figure*}

In this section, we describe how NS surface temperature $T_{s}$ is affected by the DM annihilation. If $\epsilon_{\chi}$ is negligible, the standard cooling mechanism gives $T_{s}\approx100\,{\rm K}$
for a 3-Gyr-old NS. But when $\epsilon_{\chi}$ is large enough to
counterbalance $\epsilon_{\gamma,\nu}$, $T_{s}$ could remain at
a relatively higher temperature. %
We present the numerical results of $T_{s}$ for both $\alpha_\chi=1$ and $0.01$ in Figs.~\ref{fig:Ts_ax1} and \ref{fig:Ts_ax001} respectively.
The adjacent DM density around NS is assumed to be the same as that of the solar system, $\rho_\chi=0.3 \,{\rm GeV/cm^3}$, since we aim for the nearby isolated NS.
The DM mass scale is shown from $100\,{\rm MeV}$ to $10^6\,{\rm MeV}$. Once $m_\chi\lesssim 100\,{\rm MeV}$,
all of the annihilation channels to fermions will be Pauli blocked except neutrinos.
Nonetheless, there is no upper limit for DM mass in NS. But heavier $m_\chi$ results in lesser DM number density which
makes the NS sensitivity worse. In addition, Refs.~\cite{Bramante:2017xlb,Ilie:2020vec,Dasgupta:2019juq} pointed out when $m_\chi \gtrsim \mathcal{O}(10-100)\,{\rm TeV}$, it requires multiple scatterings to capture the DM and implies that the 
single-scattering capture is inefficient.
Thus, we restrict our discussion below to the TeV DM where NS has better sensitivity and can be complementary to
current DM direct searches.

In the following, we discuss the general trends of the numerical results in terms of $\alpha_\chi=1$, 
(see Fig.~\ref{fig:Ts_ax1}) unless specified otherwise.
The values of $\sigma_{\chi n}$ and $\langle \sigma v\rangle$ are computed with the rest of the
parameters $(m_\chi,m_V,\epsilon_\gamma,\epsilon_Z)$ taking the values shown on each figure.
Constraints on these parameters according to the thermal relic density and direct searches
are displayed as well.
The conclusions can be applied to $\alpha_\chi=0.01$ directly (see Fig.~\ref{fig:Ts_ax001}).
A simple understanding on $\alpha_\chi$ is that the dark sector interactions are proportional to $\alpha_\chi^2$
and DM-SM interactions are proportional to $\alpha_\chi$. The derivations of such features on the scattering cross sections
for all interactions are given in the appendixes.

The values for the parameter $\eta\equiv \varepsilon_{\gamma}/\varepsilon_{Z}$ from
top to bottom are $1$ (combined, $\varepsilon_Z=\varepsilon_\gamma\neq 0$), 0 (pure $V-Z$ mixing, $\varepsilon_\gamma=0$)
and $\infty$ (pure kinetic mixing, $\varepsilon_Z=0$), respectively.
From left to right, we have $m_{V}/m_{\chi}=10$ (heavy mediator),
$1$ (equal mass) and $0.1$ (light mediator). $T_{s}$ is indicated
by the color bar placed on the right and the lowest temperature is $100\,{\rm K}$.
Without annihilation, e.g.~no anti-DM exists, solely kinetic heating
can raise $T_{s}$ up to $1750\,{\rm K}$. If DM annihilation is included,
$T_{s}$ can maximally reach to $3100\,{\rm K}$. 

Various constraints are also plotted, including XENON1T \cite{Aprile:2018dbl}, XENON
LDM (low mass DM) based on the ionization \cite{Aprile:2019xxb}, and of Migdal \cite{Aprile:2019jmx}
effects, SIDM \cite{Randall:2007ph,Walker:2011zu,BoylanKolchin:2011de,BoylanKolchin:2011dk,Elbert:2014bma}, SN1987A \cite{Sung:2019xie} and beam dump experiments
\cite{Riordan:1987aw,Bross:1989mp,Abdullah:2018ykz}.
The parameter curve rendering DM annihilation cross section at the 
thermal relic value \cite{Steigman:2012nb} in the early Universe
is plotted in green on each figure for comparison.\footnote{Since we have taken DM as Dirac fermions in the symmetric scenario~\cite{Lin:2011gj}, the thermal relic cross section is therefore two times larger than that in the Majorana DM case. Note that the thermal relic $\langle\sigma v\rangle$ is in general $m_\chi$ dependent and slightly deviates from the
canonical value $6\times 10^{26}\,{\rm cm^3\,s^{-1}}$ \cite{Steigman:2012nb}. We have incorporated this for determining the green curve on each figure. } 
We adopted the method given in Ref.~\cite{Cirelli:2016rnw} for computing the Sommerfeld enhancement factor.
The DM relative velocity in the early Universe is taken to be $c/3$. See Appendix \ref{sec:DM_ann} for details.
Here we present the thermal relic cross section as a reference point and refer the readers to Refs.~\cite{ArkaniHamed:2008qn,Cassel:2009wt,Lin:2011gj} for detailed discussions.
In addition, although the captured DMs can have relatively large Sommerfeld enhancement due to low velocities,\footnote{Assuming DMs are thermalized with the NS core where $T_\chi = T_b$.
	Thus the  mean velocity is about $\sqrt{T_\chi/m_\chi}$.}
the enhanced $\langle \sigma v\rangle$ only shortens the equilibrium timescale $\tau_{\rm eq}$. When $t\gg \tau_{\rm eq}$,
the total annihilation rate only depends on the capture rate with $\Gamma_A=C_c$. The NS is generally insensitive to the Sommerfeld enhancement as long as the DM is in
equilibrium.

\subsection{Case for $m_{V}\geq m_{\chi}$}

When $m_{V}\geq m_{\chi}$, only $\chi\bar{\chi}\to f\bar{f}$ is
allowed. A dip occurs on each plot in Fig.~\ref{fig:Ts_ax1} with this mass ordering. The resonant point is caused by
the pole in $\tilde{\varepsilon}_Z$ given by Eq.~(\ref{eq:Ez}) when $m_V = m_Z$ with $m_Z$ the SM $Z$ boson mass. 
In fact the value for $\tilde{\varepsilon}_Z$ at this point is $-i(\varepsilon_Z+\varepsilon_\gamma \tan\theta_W)m_Z/\Gamma_Z$, which is enhanced by the factor $m_Z/\Gamma_Z$.
 
Thus, the DM-neutron scattering cross section $\sigma_{\chi n}$ depends on $\tilde{\varepsilon}_Z$ and is proportional to
\begin{equation}
    \sigma_{\chi n} \propto  \frac{\alpha_\chi \tilde{\varepsilon}_Z^2}{m_V^4}
    \frac{m_\chi^2 m_n^2}{(m_\chi+m_n)^2}\min(\xi,1)\label{eq:sig_xn_propto}
\end{equation}
in the NR limit. (See Eq.~(\ref{eq:sigXN_nr}) for reference.)\footnote{In the numerical calculation, we used the general expression for $\sigma_{\chi n}$, Eq.~(\ref{eq:sig_XN}), and the derivation is given in the same appendix. Nonetheless, Eq.~(\ref{eq:sig_xn_propto}), or Eq.~(\ref{eq:sigXN_nr}), is simpler and suitable for our discussions in the main text.}
The last term shows the suppression factor due to Pauli blocking where $\xi\sim q/\mu_F$ with $q$ being 
the momentum transfer during the scattering and $\mu_F$ the neutron chemical potential. 

In the equilibrium epoch, $t\gg \tau_{\rm eq}$, the total annihilation
rate $\Gamma_A=C_c \propto \sigma_{\chi n}$.\footnote{We found that the equilibrium condition 
holds in most of the parameter space in this work. However, in the calculation we adopted $\Gamma_A=C_a N_{\chi}N_{\bar{\chi}}$ with $N_\chi (N_{\bar{\chi}})$ given by Eq.~(\ref{eq:Nx}), instead of simply assuming $\Gamma_A=C_c$.}
When $\tilde{\varepsilon}_Z$ is at the resonant point, $\sigma_{\chi n}$ is enhanced drastically by the factor $m_Z^2/\Gamma_Z^2$ so does the
DM heating resulted from DM emissivity $\epsilon_\chi$.
This accounts for the dip at $m_V=m_Z$ in each figure. 

On the other hand, DM heating for $m_{\chi}$ in the sub-GeV region is much stronger.
It can be understood that, as $m_\chi \ll m_n$, $q\propto m_\chi$ while $m_V/m_\chi$ is held fixed, we have
$\sigma_{\chi n}\propto \tilde{\varepsilon}_Z^2/m_\chi$ according to Eq.~(\ref{eq:sig_xn_propto}).
Hence a smaller $m_\chi$ leads to a larger $\sigma_{\chi n}$ as well as a more effective DM heating.
However, the effect of DM heating will not grow indefinitely with $\tilde{\varepsilon}_Z$ as $\sigma_{\chi n}\leq \sigma_{\chi n}^{\rm geom}\approx 10^{-44}\,{\rm cm}^2$.
The maximum $T_s$ caused by DM heating saturates when $\sigma_{\chi n}= \sigma_{\chi n}^{\rm geom}$ and
is around $3100\,{\rm K}$. This justifies our numerical results in Fig.~\ref{fig:Ts_ax1} that $T_s$
does not increase further when $\tilde{\varepsilon}_Z$ is sufficiently large for a given $m_\chi$.

For all plots in Fig.~\ref{fig:Ts_ax1}, DM heating
becomes weaker instead of proportional 
to $1/m_\chi$ for $m_\chi \lesssim \mathcal{O}(170)\,{\rm MeV}$. Although DM is capable of producing $e^\pm$
and $\mu^\pm$ in this mass range, the chemical potentials for both particles are $\mu_F^e \sim \mathcal{O}(170)\,{\rm MeV}$ and
$\mu_F^\mu \sim \mathcal{O}(70)\,{\rm MeV}$. All channels are Pauli blocked and only pions formed by 
$q\bar{q}$ are allowed until $m_\chi < m_\pi$. Nonetheless, in the presence of $V-Z$ mass mixing,
neutrinos are also part of the annihilation products and take a significant branching ratio in the DM
annihilation in such a mass region. Neutrino cannot contribute to the heating---this explains why $T_s$ is much colder when $m_\chi \lesssim \mathcal{O}(170)\,{\rm MeV}$.
The DM heating in this region is mainly due to kinetic heating.
As $\sigma_{\chi n} = \sigma_{\chi n}^{\rm geom}$, the resulted $T_s$ is around $1750\,{\rm K}$
from pure kinetic heating.

Various $\eta$ values in Fig.~\ref{fig:Ts_ax1} characterize the contributions from $\varepsilon_{\gamma,Z}$ to
$\tilde{\varepsilon}_Z$.\footnote{Since neutron charge is neutral, $Q=0$, the effect of kinetic mixing 
$Q\varepsilon_\gamma$ in Eq.~(\ref{eq:a_f}) has zero contribution to $\sigma_{\chi n}$.
Hence $\sigma_{\chi n}\propto \tilde{\varepsilon}_Z^2$. Nonetheless, if protons in the NS are considered, then
$\varepsilon_\gamma$ contributes to the DM-proton cross section $\sigma_{\chi p}$ as a 
consequence of nonvanishing $Q\varepsilon_\gamma$.}
Both $\eta=1$ and $0$ are similar because even when $\varepsilon_\gamma\neq 0$, its effect on $\tilde{\varepsilon}_Z$
is suppressed by $m_V^2/m_Z^2$ as seen from Eq.~(\ref{eq:Ez}). For $\eta=1$, the kinetic mixing can contribute
comparably to the $V-Z$ mass mixing unless $m_V> m_Z$. This can be clearly seen in Fig.~\ref{fig:Ts_ax1} that the 
difference between $\eta=1$ and $0$ is apparent only in $m_V>m_Z$ region, which is the region to the right of the dip.
To the left of the dip, the contribution from $\varepsilon_\gamma$ to $\tilde{\varepsilon}_Z$ for $\eta=1$ is
negligible.

For $\eta=\infty$, $\varepsilon_Z$ vanishes so that the only contribution to $\tilde{\varepsilon}_Z$ comes from $\varepsilon_\gamma$. As discussed earlier, the effect of kinetic mixing term is suppressed by $m_V^2/m_Z^2$ 
and thus $\sigma_{\chi n} \propto \tilde{\varepsilon}_Z^2 \propto \varepsilon_\gamma^2m_V^4/m_Z^4$.
The associated DM heating is, in general, much weaker than the cases with $\eta=1$ and $0$.
However, the advantage of $\eta=\infty$ is that no neutrinos can be produced by the DM annihilation due to
the absence of $V-Z$ mass mixing. The energy released from DM annihilation can be fully deposited into the NS. This accounts for
the higher $T_s$ than when $\eta=1$ and $0$ in terms of the same $\sigma_{\chi n}$,
but the difference is not apparent. Numerical calculation shows it is around ten to $\mathcal{O}(100)\,{\rm K}$.

\subsection{Case for $m_{V}<m_{\chi}$}

For the light mediator case, the channel $\chi\bar{\chi}\to2V$ dominates over $\chi\bar{\chi}\to f\bar{f}$ due
to $\alpha_{\chi}\gg\varepsilon_{\gamma,Z}$ in general as long as
$F\sim 1$, $V$ can fully decay into SM particles before it exits the NS.
The resulting heating from $\chi\bar{\chi}\to 2V$ with $V\to f\bar{f}$ can be appreciable as shown in
the rightmost panel of Fig.~\ref{fig:Ts_ax1}.
The heating region in the $m_V<m_\chi$ case is much more expanded than the $m_V\gg m_\chi$ case since lighter $m_V$ 
induces larger $\sigma_{\chi n}$ as shown in Eq.~(\ref{eq:sig_xn_propto}).
The resulting effects from different $\eta$'s are similar to those in the previous subsection.
When DM mainly annihilates to $2V$, the thermal relic cross section is controlled by $\alpha_\chi$ and $m_{\chi}$ while it is independent of $\varepsilon_{\gamma,Z}$.
Hence the thermal relic cross section only constrains $m_\chi$ when
$\alpha_\chi$ and $m_V$ are fixed.
For $\alpha_\chi =1$ and $0.01$, the $m_\chi$ values rendering the thermal relic cross section are around $2\times 10^8\,{\rm MeV}$ and $5\times 10^5\,{\rm MeV}$, respectively.

\section{Detectability of the future telescope }

\begin{figure}
\begin{centering}
\includegraphics[width=0.5\textwidth]{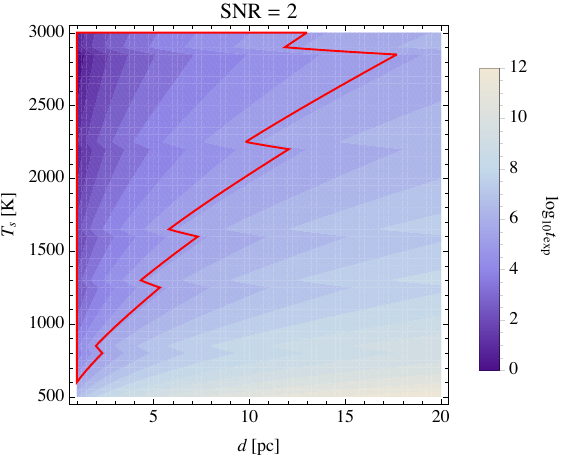}
\par\end{centering}
\caption{\label{fig:texp}The exposure time $t_{{\rm exp}}$ for ${\rm SNR}=2$
in JWST. The region enclosed by the red line indicates $t_{{\rm exp}}\protect\leq10^{5}\,{\rm s}$.}

\end{figure}

Since DM annihilation could significantly affect the NS surface temperature $T_{s}$, we
discuss the detectability of $T_s$ in the JWST and similar telescopes in the future.
The blackbody spectral flux density with $T_{s}$ at a given
frequency $\nu$ is given by \cite{Baryakhtar:2017dbj}
\begin{eqnarray}
f_{\nu}(\nu,T_{s},d)&=&\frac{4\pi^{2}\nu^{3}}{e^{2\pi\nu/k_{B}T_{s}}-1}\left(\frac{R_{0}\gamma}{d}\right)^{2} \\ \nonumber
&=&\frac{(k_BT_s)^3}{2\pi}\left(\frac{a^3}{e^a-1}\right)\left(\frac{R_{0}\gamma}{d}\right)^{2},
\end{eqnarray}
where $R_0$ is the NS radius, $\gamma\approx 1.35$ is the relativistic factor of DM on the NS surface, $d$ is the distance between the NS and the Earth, and $a=\omega/k_BT_s$. It is easy to estimate that $f_{\nu}$ peaks at $a\approx 3$ and the peak value is clearly proportional to $T_s^3$.
Taking
$T_{s}=2000\,{\rm K}$ and $d=10\,{\rm pc}$ as an example, $f_{\nu}$ peaks at $\nu^{-1}=2\,{\rm \mu m}$ with the peak value of $0.84\,{\rm nJy}$. The SNR for
JWST-like telescope scales as 
$\sqrt{t_{{\rm exp}}}$ for a given $\nu$
where $t_{{\rm exp}}$ is the exposure time. This scaling stems from the fact that, for the exposure time $t_{\rm exp}$, the signal photon number in the frequency range $[\nu,\nu+d\nu]$ is proportional to $f_{\nu}\cdot d\nu\times t_{\rm exp}$. On the other hand, the statistical fluctuation of the photon number, arising mainly from in-field and scattered zodiacal light, scattered thermal emission from the telescope, and the scattered starlight, scales as $\sqrt{t_{\rm exp}}$ for a given frequency $\nu$. This results into the above mentioned scaling for SNR. It is important to note that SNR is a complicated function of $\nu$ because the above mentioned backgrounds are also wavelength dependent~\cite{Gardner:2006ky}. 

As reported in Ref.~\cite{JWST_guide}, JWST
covers $0.8\,{\rm \mu m}$ to $5.0\,{\rm \mu m}$ imaging sensitivity
in its Near-Infrared Imager and Slitless Spectrograph (NIRISS) with multiple filters. For example the F200W filter centered at $\nu^{-1}=2\,{\rm \mu m}$
can reach ${\rm SNR}=10$ with $f_{\nu}=10\,{\rm nJy}$ and $t_{{\rm exp}}=10^{4}\,{\rm s}$. To reach the peak of $f_{\nu}$ for $T_s=2000$ K, i.e., $0.84\,{\rm nJy}$ with ${\rm SNR}=10$, one requires $1.4\times 10^6$ s of exposure time. For ${\rm SNR}=2$, which is the criterion for our presentation below, the required exposure time is $5.6\times 10^4$ s.

In Fig.~\ref{fig:texp}, we plot the $t_{{\rm exp}}$ for obtaining
${\rm SNR}=2$ over $d-T_{s}$ plane. The region enclosed by the red
line represents $t_{{\rm exp}}<10^{5}\,{\rm s}$. There are multiple
filters available for NIRI with $\nu^{-1}$ centered at various different values
\cite{JWST_guide}. We select the filter with $\nu^{-1}$ most suitably matching 
the corresponding blackbody wavelength at $T_{s}$. For instance, F200W filter is used for $T_s$ close to 2000 K, while F277W filter is adopted for $T_s$ around 1500 K.  The switching of filters when appropriate is reflected in
the zigzag behavior of the red sensitivity curve in Fig.~\ref{fig:texp}.
In principle, as $\sigma_{\chi n}\sim\sigma_{\chi n}^{{\rm geom}}$,
kinetic heating can maximally warm the NS up to $1750\,{\rm K}$ without DM annihilation.
For NS that is located within 10 pc, JWST can achieve  ${\rm SNR=2}$ with $t_{{\rm exp}}\leq10^{5}\,{\rm s}$
for $T_{s}\geq1750\,{\rm K}$

\section{Summary and outlook}

In this work we have investigated the new dynamics arising from the kinetic mixing
and $V-Z$ mass mixing between the dark gauge boson $V$ of the broken
$U(1)_{X}$ symmetry and neutral gauge bosons in SM. In particular, $V-Z$ mass mixing induces a resonance at
$m_{V}\approx m_{Z}$, which can be seen from the pole of $\tilde{\varepsilon}_{Z}$ 
at $m_{V}=m_{Z}$. The axial-vector part of the coupling between $V$ and SM fermions has been included in our calculations. As $\chi\bar{\chi}\to2V$ dominates the
annihilation channel for $m_{V}<m_{\chi}$, 
$V$ can
decay into a pair of SM fermions before it exits NS and induces NS heating in addition to $\chi\bar{\chi}\to f\bar{f}$.
Although this contribution appears naturally in the dark boson model considered here, it is usually not included in the model-independent analysis, such as the one performed in Ref.~\cite{Chen:2018ohx}.
We also demonstrated numerically that NS can provide constraints on sub-GeV DM with feeble coupling to
SM particles complementary to the current direct search.
The detectability with reasonable $t_{{\rm exp}}$
in JWST telescopes is discussed. Similar conclusion can be drawn for
the future JWST-like telescopes.

We note that this work only considers $\chi n$ scattering
in the capture rate. This explains why NS is not sensitive to the dark sector
when $\varepsilon_Z=0$ ($\eta=\infty$). Neutrons interact with DM only through NC interaction 
governed by $\tilde{\varepsilon}_{Z}$.
Once $\varepsilon_{Z}=0$, NC interaction becomes suppressed since $\varepsilon_{\gamma}$ in $\tilde{\varepsilon}_{Z}$ is oppressed by $m_V^2/m_Z^2$.
However, NS also consists of protons, although the fraction of them is rather small. 
When protons are included, charged current interaction will be involved for the capture of DM and NS remains sensitive to the dark sector even for $\varepsilon_Z=0$.
In general, NS sensitivity will be improved by including proton contributions.  
We leave this for future studies.

\appendix

\section{DM-neutron interaction\label{sec:XN_int}}
\begin{figure*}
\begin{centering}
\includegraphics{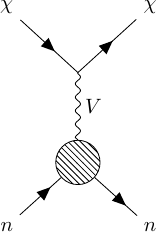}
\par\end{centering}
\caption{\label{fig:XN}Feynman diagram for DM-neutron scattering. The blob
is an effective vertex that includes both vector and axial-vector
contributions from kinetic mixing and $V-Z$ mass mixing.}

\end{figure*}
\begin{figure*}
\begin{centering}
\includegraphics{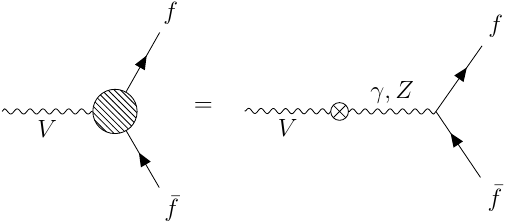}
\par\end{centering}
\caption{\label{fig:mixing_diagram}Feynman diagrams contributing to the coupling of dark boson $V$ to SM fermions. }
\end{figure*}

When DM falls into NS, they could scatter with neutrons
via exchanging the dark boson $V$ as shown in Fig.~\ref{fig:XN}. The kinetic mixing and $V-Z$ mass mixing
generate vector and axial-vector interactions between $V$ and SM fermions.
The usual derivation of these interactions proceeds through the diagonalization of both $\mathcal{L}_{\rm gauge}$ and $\mathcal{L}_{\rm mass}$ in Eqs.~(\ref{eq:L_gauge}) and (\ref{eq:L_mass}), which gives rise to relations between 
fields in the gauge basis and those in mass eigenstate basis. However, since we are only interested in interactions up to $\mathcal{O} (\varepsilon_\gamma)$ or $\mathcal{O} (\varepsilon_Z)$, we do not need to perform the diagonalization but rather treating the mixing terms 
$\varepsilon_\gamma B_{\mu\nu}V^{\mu\nu}/(2\cos\theta_W)$ and 
$\varepsilon_Z m_Z^2 Z_{\mu}V^{\mu}$ as perturbations. These two mixing terms generate the following two-point functions at the tree level
\begin{eqnarray}
i\Pi^{\mu\nu}_{V\gamma}&=&i\varepsilon_\gamma k^2g^{\mu\nu}, \nonumber \\
i\Pi^{\mu\nu}_{VZ}&=&-i(\varepsilon_\gamma\tan\theta_W k^2+\varepsilon_Zm_Z^2)g^{\mu\nu},
\end{eqnarray}	 
where $k$ is the four-momentum of $V$ entering into kinetic mixing or $V-Z$ mixing vertex. Hence the electromagnetic coupling of $V$ to SM fermions results from multiplying the two-point function $i\Pi^{\mu\nu}_{V\gamma}$, the photon propagator
$iD^\gamma_{\alpha\mu}(k)$, and the electromagnetic coupling $ieA_{\alpha}J^{\alpha}_{\rm EM}$, as shown in Fig.~\ref{fig:mixing_diagram}. 
This multiplication leads to
\begin{eqnarray}
ieJ^{\alpha}_{\rm EM} \frac{-ig_{\alpha\mu}}{k^2}i\varepsilon_\gamma k^2g^{\mu\nu}V_{\nu}=ie\varepsilon_\gamma J^{\nu}_{\rm EM}V_{\nu}.
\label{eq:EM}
\end{eqnarray}	  
Similarly, NC coupling of $V$ to SM fermions is given by multiplying the two-point function $i\Pi^{\mu\nu}_{VZ}$, the $Z$ boson propagator
$iD^Z_{\alpha\mu}(k)$, and the NC coupling $igZ_{\alpha}J^{\alpha}_{\rm NC}/\cos\theta_W$. This gives rise to 
\begin{eqnarray}
&&\frac{ig}{\cos\theta_W}J^{\alpha}_{\rm NC} \frac{-i}{k^2-m_Z^2+im_Z\Gamma_Z}\left(g_{\alpha\mu}-\frac{k_{\alpha}k_{\mu}}{m_Z^2}\right) (-i)(\varepsilon_\gamma\tan\theta_W k^2+\varepsilon_Z m_Z^2)g^{\mu\nu}V_{\nu}\nonumber \\
&=&\frac{-ig}{\cos\theta_W} J^{\nu}_{\rm NC}V_{\nu}\frac{(\varepsilon_\gamma \tan\theta_W m_V^2+\varepsilon_Z m_Z^2)}{(m_V^2-m_Z^2+im_Z\Gamma_Z)}.
\label{eq:NC}
\end{eqnarray}
Here we have used the physical conditions $k^2=m_V^2$ and $k_{\mu}\epsilon^{\mu}_{V}=0$. We have also chosen unitary gauge for the $Z$-boson propagator. 
Therefore, the interaction vertex between dark bosons and neutrons in Fig.~\ref{fig:XN} have the following Lorentz structure 
$ie\bar{\psi}_n\gamma^{\mu}(a_{f}+b_{f}\gamma^{5})\psi_n$ with
\begin{subequations}
\begin{align}
a_{f} & =Q\varepsilon_{\gamma}+\frac{1}{\sin2\theta_{W}}(I_{3}-2Q\sin^{2}\theta_{W})\tilde{\varepsilon}_{Z},\label{eq:a_f}\\
b_{f} & =-\frac{I_{3}}{\sin2\theta_{W}}\tilde{\varepsilon}_{Z},\label{eq:b_f}
\end{align}
\end{subequations}
where
\begin{equation}
\tilde{\varepsilon}_{Z}=\frac{\varepsilon_{Z}+\varepsilon_{\gamma}\tan\theta_{W} (m_{V}^{2}/m_{Z}^{2})}{(1-m_{V}^{2}/m_{Z}^{2})^{2}+\Gamma^2_{Z}/m_Z^2}\left(1-\frac{m_{V}^{2}}{m_{Z}^{2}}-i\frac{\Gamma_{Z}}{m_Z}\right)\label{eq:Ez}
\end{equation}
and $\Gamma_{Z}$ is the $Z$ boson decay width, $Q$ and $I_{3}$
are the electric charge and the weak isospin respectively. In Table \ref{tab:a_b}, we list $Q$ and $I_{3}$
for various particles. The values for neutrons can be obtained by summing
the corresponding quantum numbers of three quarks $udd$ in the low energy limit.

Mixing parameters $\varepsilon_{\gamma}$ and $\tilde{\varepsilon}_{Z}$ are
responsible for electromagnetic and NC interactions, respectively.
Electromagnetic interaction does not contribute to $\sigma_{\chi n}$
since $Q=0$ for neutrons. 
On the other hand $\tilde{\varepsilon}_{Z}$ has a feeble dependence on $\varepsilon_{\gamma}$
with a suppression factor $m_{V}^{2}/m_{Z}^{2}$ when $m_V\ll m_Z$. This explains why
$\sigma_{\chi n}$ is still nonzero when $\varepsilon_Z=0$ ($\eta=\infty$).

\begin{table*}
\begin{centering}
\begin{tabular}{ccccccccc}
\hline 
 & $u$ & $d$ & $c$ & $s$ & $t$ & $b$ & $\ell$ & $\nu$\tabularnewline
\hline 
\hline 
$Q$ & $\frac{2}{3}$ & $-\frac{1}{3}$ & $\frac{2}{3}$ & $-\frac{1}{3}$ & $\frac{2}{3}$ & $-\frac{1}{3}$ & $-1$ & $0$\tabularnewline
$I_{3}$ & $\frac{1}{2}$ & $-\frac{1}{2}$ & $\frac{1}{2}$ & $-\frac{1}{2}$ & $\frac{1}{2}$ & $-\frac{1}{2}$ & $-\frac{1}{2}$ & $\frac{1}{2}$\tabularnewline
\hline 
\end{tabular}
\par\end{centering}
\caption{\label{tab:a_b}Values of $Q$ and $I_{3}$ for quarks,
leptons and neutrinos.}
\end{table*}

The spin-averaged $\chi n$ scattering amplitude is given by
\begin{align}
\overline{|\mathcal{M}_{\chi n}|^{2}} & =\frac{8\pi\alpha_{\chi}}{(t-m_{V}^{2})^{2}}\{-4m_{n}^{2}[(b_{f}^{2}-a_{f}^{2})m_{\chi}^{2}+a_{f}^{2}u+b_{f}^{2}(s+u)]+2(a_{f}^{2}+3b_{f}^{2})m_{\chi}^{4}\nonumber \\
 & \quad-4a_{f}^{2}um_{n}^{2}+a_{f}^{2}(t^{2}+2tu+2u^{2})+2(a_{f}^{2}-b_{f}^{2})m_{n}^{4}+b_{f}^{2}(s^{2}+u^{2})\},\label{eq:XN_amp}
\end{align}
where $s$, $t$, and $u$ are the Mandelstam variables.
DM scatters with neutrons when its velocity boosted to $0.3c-0.6c$ by
the NS gravity. It must be treated relativistically.
However,
neutrons can be treated as at rest since the chemical potential is $\mathcal{O}(200)\,{\rm MeV}$ in the star.

Therefore, from the method in Ref.~\cite{Ilisie:2016jta}, we are able to write
down the DM-neutron scattering cross section as
\begin{equation}
\sigma_{\chi n}=\frac{1}{16\pi\lambda^{1/2}(s,m_{1}^{2},m_{2}^{2})\lambda^{1/2}(s,m_{3}^{2},m_{4}^{2})}\int_{t_{-}}^{t_{+}}\overline{|\mathcal{M}_{\chi n}|^{2}}dt\label{eq:sig_XN}
\end{equation}
where
\begin{equation}
\lambda(x,y,z)=x^{2}+y^{2}+z^{2}-2xy-2yz+2xz
\end{equation}
is the K\"{a}ll\'{e}n function,
\begin{equation}
t_{\pm}=\frac{1}{2}\sum_{i=1}^{4}m_{i}^{2}-\frac{s}{2}-\frac{1}{2s}(m_{1}^{2}-m_{2}^{2})(m_{3}^{2}-m_{4}^{2})\pm\frac{\lambda^{1/2}(s,m_{1}^{2},m_{2}^{2})\lambda^{1/2}(s,m_{3}^{2},m_{4}^{2})}{2s},
\end{equation}
and
\begin{equation}
s=m_{1}^{2}+m_{2}^{2}+2E_{1}m_{2},\label{eq:s}
\end{equation}
where $m_{1}=m_{3}=m_{\chi}$ and $m_{2}=m_{4}=m_{n}$ for the $\chi n$
scattering. In Eq.~(\ref{eq:s}), the energy $E_{1}=\gamma m_{1}$
is the total energy carried by particle one, which is DM. 

\subsection{Pauli blocking in the $\chi n$ scattering}

Note that if the momentum transfer $\sqrt{-t}$ in Eq.~(\ref{eq:sig_XN}) is smaller than the Fermi momentum, the suppression by Pauli blocking takes effect. We include this in the numerical calculation by incorporating the method in Ref.~\cite{Bell:2020jou}. Our
 result agrees with Ref.~\cite{Bell:2020jou} in the three benchmark
scenarios that $\overline{|\mathcal{M}_{\chi n}|^{2}}$ are constant,
$t$-dependent and $t^{2}$-dependent.

\subsection{Axial-vector contribution in the NR limit}

If $\chi$ can be treated nonrelativistically as well, we have $s=m_{\chi}^{2}+m_{n}^{2}+2m_{\chi}m_{n}$,
$u=m_{\chi}^{2}+m_{n}^{2}-2m_{\chi}m_{n}$ and $t=0$. Therefore the
amplitude and the cross section become,
\begin{equation}
\overline{|\mathcal{M}_{\chi n}^{{\rm NR}}|^{2}}=\frac{64\pi a_{f}^{2}\alpha_{\chi}m_{\chi}^{2}m_{n}^{2}}{m_{V}^{4}}
\end{equation}
and 
\begin{equation}
\sigma_{\chi n}^{{\rm NR}}=\frac{4a_{f}^{2}\alpha_{\chi}}{m_{V}^{4}}\frac{m_{\chi}^{2}m_{n}^{2}}{(m_{\chi}+m_{n})^{2}}\label{eq:sigXN_nr}
\end{equation}
which are independent of $b_f$ where it determines the strength of axial-vector coupling.

\section{DM annihilation}\label{sec:DM_ann}

\begin{figure*}
\begin{centering}
\subfloat[\label{fig:XX2ff}To SM particles]{\begin{centering}
\includegraphics{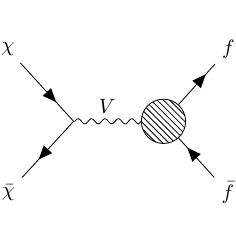}
\par\end{centering}

}\qquad\subfloat[\label{fig:XX2VV}To dark bosons]{\begin{centering}
\includegraphics{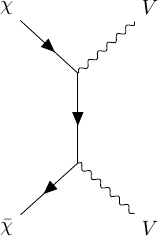}
\par\end{centering}
\quad\includegraphics{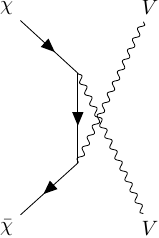}}
\par\end{centering}
\caption{Various channels for DM annihilation}
\end{figure*}

We can divide the DM annihilation into two categories, which are 
$m_{V}\geq m_{\chi}$ and  $m_{V}<m_{\chi}$, respectively. For the
prior case, DM can only annihilate into SM particles as shown in Fig.~\ref{fig:XX2ff}.
For the later one, as long as $g_{d}\gg e\varepsilon_{\gamma,Z}$,
the dominant annihilation products are two dark bosons $V$ as shown in Fig.~\ref{fig:XX2VV}.
The amplitude for $\chi\bar{\chi}\to f\bar{f}$ is given by
\begin{align}
\overline{|\mathcal{M}_{\chi\bar{\chi}\to f\bar{f}}|^{2}} & =\frac{8\pi \alpha_\chi}{(s-m_{V}^{2})^{2}+m_{V}^{2}\Gamma_{V}^{2}}\{a_{f}^{2}[-2(m_{f}^{2}+m_{\chi}^{2})(-m_{f}^{2}-m_{\chi}^{2}+2u)+s^{2}+2su+2u^{2}]\nonumber \\
 & \quad+b_{f}^{2}[-4m_{\chi}^{2}(m_{f}^{2}+t+u)-2m_{f}^{4}+6m_{\chi}^{4}+t^{2}+u^{2}]\}
\end{align}
where $\Gamma_V$ is the $V$ decay width. 
Assuming DM is at rest in the star, the amplitude can be simplified into
\begin{equation}
\overline{|\mathcal{M}_{\chi\bar{\chi}\to f\bar{f}}|^{2}}=\frac{128\pi\alpha_{\chi}}{(4m_{\chi}^{2}-m_{V}^{2})^{2}+m_{V}^{2}\Gamma_{V}^{2}}m_{\chi}^{4}\left[a_{f}^{2}\left(1+\frac{1}{2}\frac{m_{f}^{2}}{m_{\chi}^{2}}\right)+b_{f}^{2}\left(1-\frac{m_{f}^{2}}{m_{\chi}^{2}}\right)\right].\label{eq:NR_M_XX2ff}
\end{equation}
The partial decay widths of $V$ are given by 
\begin{equation}
\Gamma_{V}=\frac{m_{V}}{12\pi}\sqrt{1-4\frac{m_{f}^{2}}{m_{V}^{2}}}\left[a_{f}^{2}\left(1+2\frac{m_{f}^{2}}{m_{V}^{2}}\right)+b_{f}^{2}\left(1-4\frac{m_{f}^{2}}{m_{V}^{2}}\right)\right]\label{eq:V2ff}
\end{equation}
for $V\to f\bar{f}$ and
\begin{equation}
\Gamma_{V}=\frac{\alpha_{\chi}}{3}m_{V}\sqrt{1-4\frac{m_{\chi}^{2}}{m_{V}^{2}}}\left(1+2\frac{m_{\chi}^{2}}{m_{V}^{2}}\right)\label{eq:V2xx}
\end{equation}
for $V\to\chi\bar{\chi}$. Note that we have omitted the Heaviside theta
function $\theta(m_{V}-2m_{\chi,f})$ in the above expressions but
it is always implemented when we perform the calculation to ensure the energy
conservation. Besides, when $m_{\chi}>m_{V}$, the channel $\chi\bar{\chi}\to2V$
is allowed and the amplitude is 
\begin{align}
\overline{|\mathcal{M}_{\chi\bar{\chi}\to2V}|^{2}} & =-\frac{32\pi^2\alpha_\chi{2}}{(t-m_{\chi}^{2})^{2}(u-m_{\chi}^{2})^{2}}\{m_{V}^{4}[6m_{\chi}^{2}(t+u)-6m_{\chi}^{4}+t^{2}-8tu+u^{2}]\nonumber \\
 & \quad+4m_{V}^{2}[m_{\chi}^{4}(t+u)-4m_{\chi}^{2}tu+tu(t+u)]-m_{\chi}^{4}(3t^{2}+14tu+3u^{2})\nonumber \\
 & \quad+m_{\chi}^{2}(t^{3}+7t^{2}u+7tu^{2}+u^{3})+6m_{\chi}^{2}-tu(t^{2}+u^{2})\}.\label{eq:M2xxVV}
\end{align}
In the NR limit,
\begin{equation}
\overline{|\mathcal{M}_{\chi\bar{\chi}\to2V}|^{2}}=256\pi^{2}\alpha_{\chi}^{2}\frac{m_{\chi}^{2}(m_{\chi}^{2}-m_{V}^{2})}{(m_{V}^{2}-2m_{\chi}^{2})^{2}}.\label{eq:NR_M_XX2VV}
\end{equation}
Thus, the general expression for annihilation cross section is obtained by using the Fermi golden
rule,
\begin{equation}
\sigma v=\frac{\overline{|\mathcal{M}|^{2}}}{32\pi m_{\chi}^{2}}\sqrt{1-\frac{m_{f}^{2}}{m_{\chi}^{2}}}\theta(m_{\chi}-\mu_{F}^{f}),
\end{equation}
where $m_{f}$ is the final state particle mass and $\mu_{F}^{f}$
the chemical potential of fermion $f$ in the star. There is no chemical potential for dark boson $V$.
Therefore, we arrive at
\begin{equation}
(\sigma v)^{f\bar{f}}=4\alpha_{\chi}\kappa m_{\chi}^{2}\sqrt{1-\frac{m_{f}^{2}}{m_{\chi}^{2}}}\theta(m_{\chi}-\mu_{F}^{f}),
\end{equation}
where
\begin{equation}
\kappa=\frac{1}{(4m_{\chi}^{2}-m_{V}^{2})^{2}+m_{V}^{2}\Gamma_{V}^{2}}\left[a_{f}^{2}\left(1+\frac{1}{2}\frac{m_{f}^{2}}{m_{\chi}^{2}}\right)+b_{f}^{2}\left(1-\frac{m_{f}^{2}}{m_{\chi}^{2}}\right)\right]
\end{equation}
for $\chi\bar{\chi}\to f\bar{f}$ and
\begin{equation}
(\sigma v)^{2V}=8\pi\alpha_{\chi}^{2}\sqrt{1-\frac{m_{V}^{2}}{m_{\chi}^{2}}}\frac{(m_{\chi}^{2}-m_{V}^{2})}{(m_{V}^{2}-2m_{\chi}^{2})^{2}}
\end{equation}
for $\chi\bar{\chi}\to2V$. The total annihilation cross section is
the sum of both
\begin{equation}
\sigma v=(\sigma v)^{f\bar{f}}+(\sigma v)^{2V}.\label{eq:sigv_tot}
\end{equation}
We note that the second term contributes when $m_{V}<m_{\chi}$.

\section{Dark boson in the star\label{sec:V_in_NS}}

Dark bosons can be produced from DM annihilation once $m_{V}<m_{\chi}$.
This channel is thought to have feeble effect on the heating since
$V$ interacts weakly with the NS medium and escapes without any trace.
However, we found that, depending on the strength of $\varepsilon_{\gamma,Z}$,
$V$ can decay into SM particles before it reaches the surface of the star.
In the case that the decay length $\ell_{{\rm dec}}$
is much smaller than the radius of the star, the total energy released
from the annihilation can be fully deposited to the star. See Fig.~\ref{fig:V_decay}.
We also examine the case where $V$ is produced in the DM rich region
in the center of the star. $V$ could undergo multiple scattering with the
surrounding DM and self-trapped until it decays. See Fig.~\ref{fig:V_self_trap}. This
is another way to extract energy from $V$.
We discuss both effects in the following.

\subsection{Decay length\label{subsec:decay_len}}

\begin{figure*}
\begin{centering}
\includegraphics[width=0.9\textwidth]{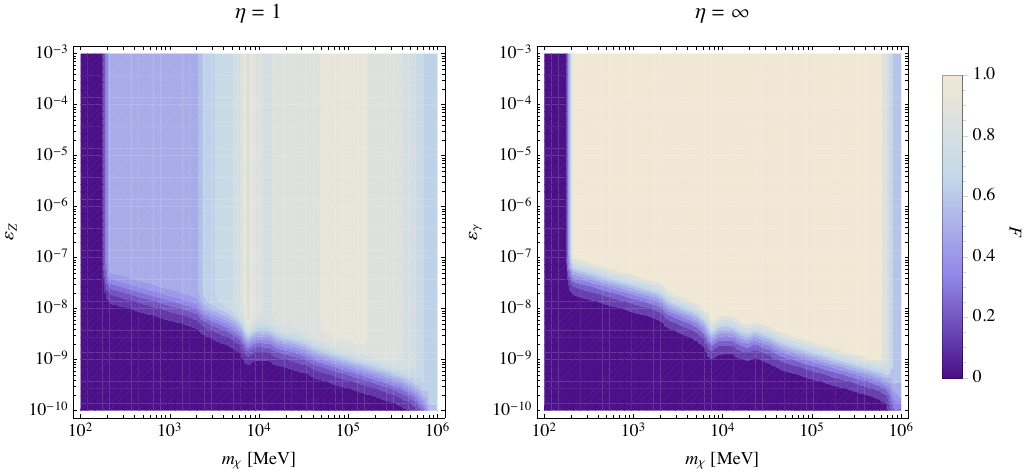}
\par\end{centering}
\caption{\label{fig:F}Fraction $F$ of dark boson decay into SM particles
that contribute to the heating effect.}

\end{figure*}

The dark boson decay length with time dilation effect is given by
\begin{equation}
\ell_{{\rm dec}}=v\gamma\tau_{{\rm dec}},
\end{equation}
where $v=\sqrt{1-m_{V}^{2}/m_{\chi}^{2}}$ is the $V$ velocity and
$\tau_{{\rm dec}}=\Gamma_{V}^{-1}$ the $V$ lifetime at rest. Let us assume that
$V$ is produced in the center of the star and its propagation distance is $R_0$.
Fig.~\ref{fig:F} presents $F$ defined in Eq.~(\ref{eq:F}), i.e., the fraction of $V$ converting into SM particles 
after traveling a distance $r=R_0$, as functions of $\varepsilon_{Z,\gamma}$ and $m_{\chi}$ for $m_{V}=0.1m_{\chi}$.
We have subtracted neutrino contributions from $F$ since they
cannot generate heat. Since the branching ratio of $V$ decays to neutrinos is nonzero in the case of
$V-Z$ mass mixing, $F$ is generally smaller than $1$ for $\eta=1$. For $\eta=\infty$, no neutrinos can be produced, thus
$F$ can reach unity.

In these figures, the chemical potential for electron $\mu_{F}^{e}$
is about $\mathcal{O}(170)\,{\rm MeV}$. For a dark boson at rest with $m_{V}\leq\mu_{F}^{e}$, $V\to e^{+}e^{-}$ can be Pauli blocked even for $m_{V}\geq2m_{e}$.
On the other hand, if $V$ is highly boosted as a result of heavy DM annihilation, $V\to e^+e^-$ is not Pauli blocked as long as  $m_{\chi}\geq\mu_{F}^{e}$. 
Therefore, to enable $V\to f\bar{f}$ decays,
two conditions are required. The first is $m_{\chi}\geq\mu_{F}^{f}$
and the second is $m_{V}\geq2m_{f}$.


\subsection{Dark boson-DM interaction length\label{subsec:VX_scatt_len}}

\begin{figure*}
\begin{centering}
\includegraphics{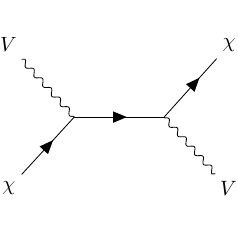}\qquad\includegraphics{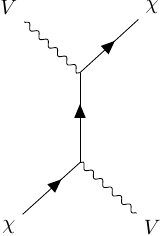}
\par\end{centering}
\caption{\label{fig:XV}$\chi V$ scattering via $s$ and $t$ channels.}
\end{figure*}

Feynman diagrams contributing to $\chi V$ scattering are shown in Fig.~\ref{fig:XV}
and the amplitude is given by
\begin{align}
\overline{|\mathcal{M}_{\chi V}|^{2}} & =\frac{64\pi^{2}}{3}\frac{\alpha_{\chi}^{2}}{(s-m_{\chi}^{2})^{2}(t-m_{\chi}^{2})^{2}}\{m_{V}^{4}[6m_{\chi}^{2}(s+t)-6m_{\chi}^{4}+s^{2}-8st+t^{2}]\nonumber \\
 & \quad-m_{\chi}^{4}(3s^{2}+14st+3t^{2})+m_{\chi}^{2}(s^{3}+7s^{2}t+7st^{2}+t^{3})\nonumber \\
 & \quad+4m_{V}^{2}[m_{\chi}^{4}(s+t)-4stm_{\chi}^{2}+st(s+t)]+6m_{\chi}^{8}-st(s^{2}+t^{2})\}.\label{eq:XV}
\end{align}
To compute the scattering cross section $\sigma_{\chi V}$, it is fair to assume
DM at rest. However, $V$ is produced with relativistic velocity since $m_\chi > m_V$.
We follow the procedure given in Eqs.~(\ref{eq:sig_XN})-(\ref{eq:s})
and set $m_{1}=m_{3}=m_{V}$ and $m_{2}=m_{4}=m_{\chi}$. Thus,
\begin{equation}
\sigma_{\chi V}=\frac{1}{16\pi\lambda(s,m_{V}^{2},m_{\chi}^{2})}\int_{t_{-}}^{t_{+}}\overline{|\mathcal{M}_{\chi V}|^{2}}dt.\label{eq:sig_XV}
\end{equation}
Note that $\chi V$ scattering is not subject to Pauli blocking since
DM does not become degenerate in the presence of annihilation.

\begin{figure*}
\begin{centering}
\includegraphics[width=0.9\textwidth]{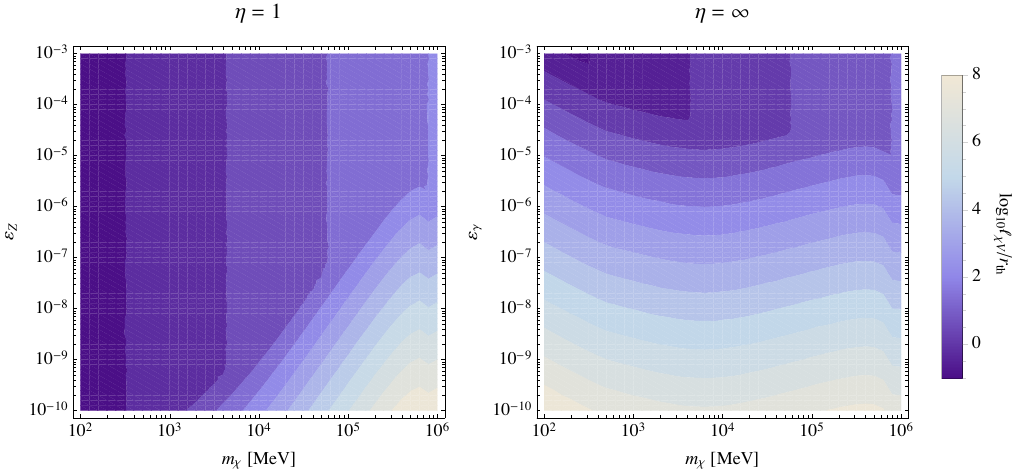}
\par\end{centering}
\caption{\label{fig:l2r}The ratio $\ell_{\chi V}/r_{{\rm th}}$ for $\eta=1$
and $\infty$. We take $\alpha_{\chi}=1$ and $T_{\chi}=1000\,{\rm K}$
in the calculation.}

\end{figure*}

The $\chi V$ scattering length $\ell_{\chi V}$ is given by 
\begin{equation}
\ell_{\chi V}=(n_{\chi}\sigma_{\chi V})^{-1},
\end{equation}
with $n_{\chi}\equiv N_{\chi}/V_{\chi}$ the average DM number density.
The volume characterizing DM in NS is $V_{\chi}=4\pi r_{{\rm th}}^{3}/3$, where 
\begin{equation}
r_{{\rm th}}\approx2.4\times10^{3}\,{\rm cm}\,\left(\frac{T_{\chi}}{10^{5}\,{\rm K}}\frac{10\,{\rm MeV}}{m_{\chi}}\right)^{1/2}\label{eq:rth}
\end{equation}
is the thermal radius.
If $\ell_{\chi V}\ll r_{{\rm th}}$, $V$ can scatter with surrounding DM multiple times and gradually lose its kinetic energy.
However, our numerical result shows that $\ell_{\chi V}\gg r_{\rm th}$ in all of 
our interested parameter space.
In Fig.~\ref{fig:l2r}, we take $\alpha_{\chi}=1$ and $T_{\chi}=1000\,{\rm K}$.
The choice $\alpha_{\chi}<1$ makes $\ell_{\chi V}$ even longer due
to a weaker $\chi V$ interaction.
For $\eta=1$, the region for $\ell_{\chi V}/r_{{\rm th}}<1$ happens
when $m_{\chi}\lesssim300\,{\rm MeV}.$ However, even $V$ can be
self-trapped, it hardly decays into particles other than neutrinos
because the allowed channels, eg.~$e^{\pm}$ and $\mu^{\pm}$, are
Pauli blocked. For $\eta=\infty$, only a very small parameter space
leads to $\ell_{\chi V}/r_{{\rm th}}<1$.
Therefore, we conclude that the self-trapping of $V$ 
is insignificant, hence only $V$ decays contribute to the energy injection.



\begin{acknowledgments}
G.~L.~L.~is supported by the Ministry
of Science and Technology, Taiwan under Grant No.~107-2119-M-009-017-MY3. Y.~H.~L.~is supported by the Postdoctoral Scholar Program of the Academia Sinica, Taiwan.
\end{acknowledgments}

\end{document}